\documentclass[final,3p,times]{elsarticle}

\usepackage{graphicx}
\usepackage{amssymb}

\usepackage{lineno}
\usepackage{graphicx}
\usepackage{float}
\usepackage{longtable,multicol,relsize }
\usepackage{epsfig,amsfonts,amsmath}
\usepackage{bm}

\usepackage{caption}
\usepackage{subcaption}
\usepackage{amsmath}
\usepackage{wasysym}
\usepackage{bigints}
\usepackage{diagbox}
\usepackage{todonotes}
\usepackage{xcolor}

\usepackage{lineno}
\usepackage{epsfig,amsfonts,amsmath}
\usepackage{wrapfig}
\usepackage{breqn,soul}
\def\correspondingauthor{\footnote{Corresponding author.}}

\newcommand{\vect}[1]{\boldsymbol{\mathit{#1}}}

\newcommand{\adj}{\mathrm{adj}}

\renewcommand{\det}{\mathrm{det}}

\newcommand{\tens}[1]{\mathbf{#1}}

\newcommand{\tenf}[1]{{\mathbb{#1}}}%

\journal{Polymers}

\begin{document}

\begin{frontmatter}


\title{A Physics-informed Assembly of Feed-Forward Neural Network Engines to Predict Inelasticity in Cross-Linked Polymers}

\author[label1,label2]{Aref Ghaderi}
\author[label1,label3]{Vahid Morovati}

\author[label1,label4]{Roozbeh Dargazany\correspondingauthor{}}

\address[label1]{Department of Civil and Environmental Engineering, Michigan State University}
\address[label2]{ghaderi1@msu.edu}
\address[label3]{morovati@msu.edu}
\address[label4]{roozbeh@msu.edu}

\begin{abstract}
In solid mechanics, data-driven approaches are widely considered as the new paradigm that can overcome the classic problems of constitutive models such as limiting hypothesis, complexity, and accuracy.
However, the implementation of machine-learned approaches in material modeling has been modest due to the high-dimensionality of the data space, the significant size of missing data, and limited convergence. This work proposes a framework to hire concepts from polymer science, statistical physics, and continuum mechanics to provide super-constrained machine-learning techniques of reduced-order to partly overcome the existing difficulties. Using a sequential order-reduction, we have simplified the 3D stress-strain tensor mapping problem into a limited number of super-constrained 1D mapping problems. Next, we introduce an assembly of multiple replicated Neural Network learning agents (L-agents) to systematically classify those mapping problems into a few categories, each of which were described by a distinct agent type.
By capturing all loading modes through a simplified set of dispersed experimental data, the proposed hybrid assembly of L-agents provides a new generation of machine-learned approaches that simply outperform most constitutive laws in training speed, and accuracy even in complicated loading scenarios. Interestingly, the physics-based nature of the proposed model avoids the low interpretability of conventional machine-learned models.

\end{abstract}


\end{frontmatter}

\section{\label{intro}Introduction}

\begin{figure}[H]
\centerline{\includegraphics[width=.9\textwidth]{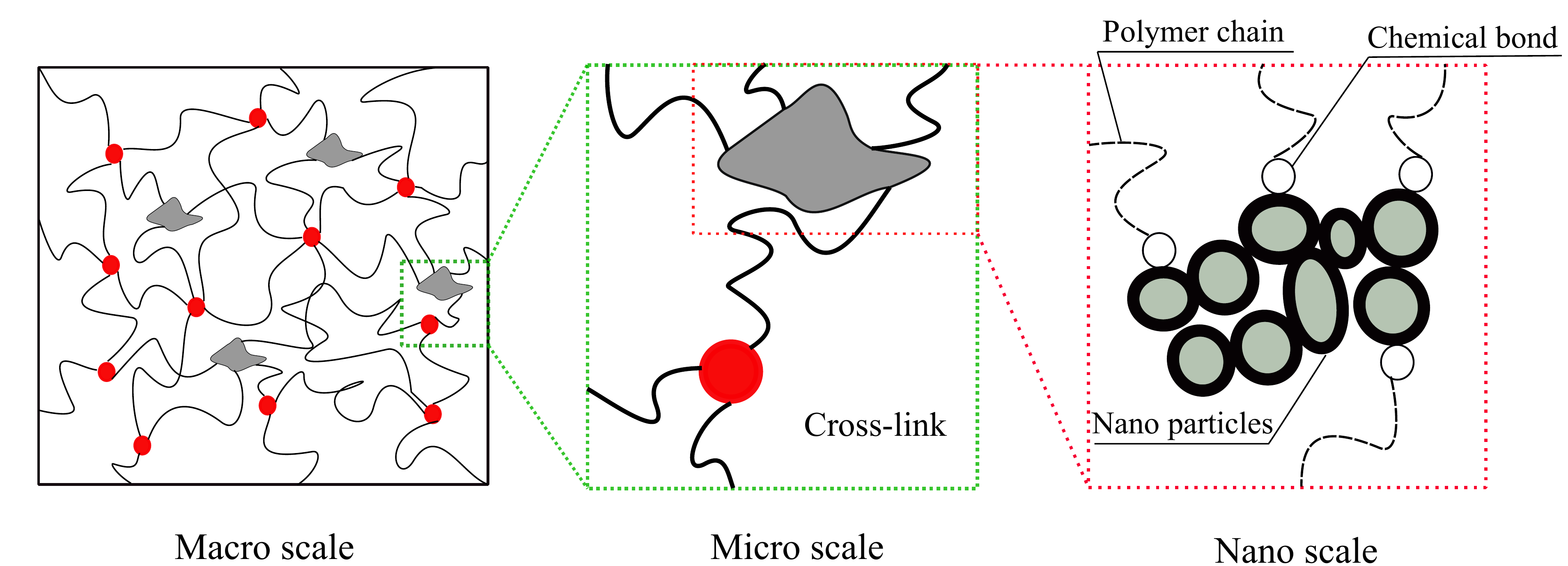}}
\caption{Schematic illustration for micro-structure of filled elastomers which is depicted micro scale and nano scale}
\label{polymer}
\end{figure}

The wide range applications of cross-linked polymers in several industries such as automotive, structural, medical, to name but a few, have made them an attractive area of research. These materials have a 3D network configuration consisting of randomly oriented long molecular chains, which are cross-linked, spiraled, and tangled among themselves or neighbors. They are typically classified into filled and unfilled categories. Fillers, in most cases, can reinforce polymers (see Fig. \ref{polymer}). Regarding the various applications of these materials, modeling their mechanical behavior in a broad range of strains is of great importance. In quasi-static deformations, these materials show hyper-elastic behavior. This behavior is dominantly governed by changes in network entropy, where the chains reorient in response to the applied macroscopic deformations. Farhangi et al. investigated effect of fiber reinforced polymer tubes filled with recycled materials \cite{farhangi12020effect, farhangi2020effect}. Izadi et al. investigated effect of nanoparticles on mechanical properties of polymers \cite{izadi2019plasma, sinha2019novel, izadi2020mechanical}. Accordingly, several studies have investigated this hyper-elastic behavior based on phenomenological or micromechanical approaches which use statistics of molecular chains network (see reviews \cite{marckmann2006comparison, steinmann2012hyperelastic}). Shojaeifar et al. \cite{shojaeifard2020finite, shojaeifard2020rutting} developed a model for modeling of visco-hyperelasticity of materials.  Phenomenological approaches are empirical, simple, and less interpretable; however, micromechanical approaches are highly interpretable but complex because they consider the readjustment of kinks, the rearrangement of convolutions, reorientation, and uncoiling of molecular chains. Meanwhile, the emergence of machine-learned (ML) models has attracted much attention as a way to address the mentioned challenges of the phenomenological and micromechanical approaches.

The exponential growth of computational power over the last decade has enabled the first-generation of ML models to be used in computational mechanics and polymer physics \cite{liu2019computational, montans2019data, lu2019data, chen2019application}. Current ML models were often developed based on "black box" approaches, which besides low interpretability, require a large volume of training data to prescribe a particular behavior \cite{vahidi2020memory, tamhidi2020conditioned}. In solid mechanics, stress-strain tensors are only partially observable in lower-dimensions. Thus, obtaining data to feed the black-box ML model is exceptionally challenging. In general, one can classify current Data-driven efforts in computational mechanics into three categories with lots of approaches placed between two categories (see review \cite{bock2019review}).

    \textbf{- Model-free Distance-minimization Approaches} were developed to circumvent the need for constitutive models by directly finding stress-strain pairs with the least distance to experimental data, which also satisfy compatibility and equilibrium constraints. This approach was initially set for nonlinear truss and linear elastic materials \cite{kirchdoerfer2016data} and later were expanded to include hyper-elastic materials \cite{nguyen2018data}. While being superior to other models by being statistically independent of any prior knowledge of the materials, the method has few major limitations. It has a excessively high computational cost, has strong sensitivity to data scattering, and in high-dimensional problems suffers from lack of data \cite{leygue2018data}. This approach is further amended by studies on the combination of data-driven identification and computational mechanics \cite{stainier2019model}. To reformulate the heuristic optimization approach adopted by \cite{kirchdoerfer2016data}, mixed-integer programming was used for its implementation \cite{kanno2019mixed}. Coelho and Breitkopf \cite{breitkopf2013multidisciplinary} in their book investigated the main approaches for constitutive modeling using optimization methods.

    \textbf {- Non-linear Dimensionality Reduction Approaches} seek to build a constitutive manifold from experimental data to describe an accurate approximation of the strain energy in different states of deformation. These approaches focus on describing the constitutive behavior through a set of shape functions, such as $B-spline$ \cite{amores2019average}, with constants derived through the LSQ error minimization \cite{ibanez2018manifold, ibanez2017data, amores2020data} or a ML approach \cite{ibanez2019hybrid}. Mainly derived from the WYPiWYG model \cite{latorre2017wypiwyg}, it focuses on solving the system of linear equations which consist coefficients of shape functions, rather than nonlinear fitting a predefined model. In elasticity, manifold learning is more efficient and more accurate than black-box ML models and it has already been generalized to cover damage \cite{minano2018wypiwyg}. In Matous's study, a manifold-based reduced order model was proposed \cite{bhattacharjee2016nonlinear}. This model relies on non-linear dimensionality reduction and the connection of macroscopic loading parameters to reduced space using an artificial neural network (ANN). Fritzen et. al \cite{fritzen2018two} proposed a data-driven homogenization method for hyper-elastic solids using the reduced order method. In their work, the surrogate model combines radial basis functions and piece-wise cubic polynomials. The main problem with these approaches is the large number of tests needed for validation and their dependency on the assumption of constitutive manifolds with a particular functional structure \cite{amores2020data}.

    \textbf{- Autonomous Approaches}
    incorporate ML models as surrogate functions to capture the high-dimensional and non-smooth micro-scale behavior of material constituents, which has been shown to be a successful approach in Multi-scale analysis \cite{reimann2019modeling}. Several multi-scale methods of analysis have been proposed based on the implementation of micro-scale ML models into the reduced-order FE simulations of the macro-scale approach \cite{lu2019data}. This coupling allows for the scalable utilization of ML surrogate models. However, the validity range of current ML models is extremely limited due a number of reasons (i) the unconstrained search space of optimization variables, (ii) neglecting underlying physics, (iii) difficulties in deriving parameter feasibility ranges, and (iv)  lack of transition models to reduce the order of the problem. Recently by implementing the reinforcement learning concept, a new class of ML meta-models have been successfully developed based on (non)cooperative games, where the model trains a pair of L-agents to emulate a specific performance through turn-based trial and error \cite{wang2019cooperative}. This paradigm employs ML techniques to capture the behavior and interaction of microstructures as a surrogate model. In Stoffel's study \cite{stoffel2019neural}, they replaces the viscoplastic material law in finite element simulation with a feed-forward neural network to make an intelligent element. Another study \cite{kopal2018prediction} was conducted to predict the tension response of rubber by a feed-forward neural network. They used strain values and filler percentages as inputs generated from a regression model and stress as output. Kaliske and Zopf \cite{zopf2017numerical} proposed an inelastic model-free approach represented by recurrent neural networks for uncured elastomers. For history-dependent functions, naturally, recurrent neural networks offer attractive alternatives, but require enormous amounts of training paths of standardized lengths, which are highly non-trivial. In 2019, Haghighat et al. \cite{raissi2019physics} proposed a physics-informed neural network that solves any given law of physics described by non-linear partial differential equations. Another study \cite{haghighat2020deep} showed that the performance of this model for linear elasticity, and Xu et al. \cite{xu2020inverse} modeled viscoelastic materials using physic constrained learning. Recently, we developed a a Bayesian surrogate constitutive model based on Bayesian regression and Gaussian process \cite{ghaderi2020bayesian} to consider uncertainty of model \cite{tooranjipour2019prescribed}. A recent study \cite{lu2019data} proposed a data-driven constitutive model by predicting a non-linear constitutive law using a neural network surrogate model constructed using a learning phase on a set of RVE non-linear computations. An investigation was conducted to formulate a constitutive model for rate-dependent materials by neural network and its implementation in finite element analysis. The challenge of sufficient data set for training, however, still remains \cite{jung2006neural}.

 Here, a cooperative multi-agent system $\mathcal B^{ {\vect{d}_{i}}}=\mathcal{A}^i_j, \,\, i \in \left\{1,n\right\},\,\, j\in \left\{1,m\right\}$ is proposed to describe different features in material behavior by using $n \times m$  different machine-learned agents (L-agents) which learn from experimental data sets. To reduce problem dimensionality, the 3D matrix is represented by $m$ 1D directions, which allows researchers to replicate each L-agent $m$ times to represent the 1D behavior of the material. The proposed model trains each agent to emulate a certain material behavior with the objective function being the error between the overall prediction of the system and the experimental data. Model fusion is used to integrate all L-agents back into a centralized system.
 
 The main contributions of this work is to infuse knowledge of physics into the model through certain modeling constraints, namely (1) by providing a new data-driven model based on physics behind a machine learning process for predicting non-linear mechanical behavior of cross-linked polymers (2) the first data-driven model that captures inelastic behavior of cross-linked polymers such as Mullins effect and permanent set (3) a new paradigm with the upgrade-ability of model from hyper-elastic to damage behavior roots from easy transformation from the integration of micro-mechanics to the machine learning process (4) proposing a new model with better training speed and accuracy compared to several well-known models. {There are two types of cross-linked polymers. One type shows hyper-elastic behavior; however, another type does not have hyper-elasticity at all. In this study, our focus is on cross-linked polymers which have hyper-elasticity such as rubbers and elastomers.} This paper is organized as follows; in section \ref{CM}, the main concepts of non-linear behavior and deformation-induced damage in cross-linked polymers are introduced and described in detail. Section \ref{PDR} explains the idea and formulation of the proposed model in detail. Model verification with experimental data on rubber inelasticity is discussed in section \ref{IRI}. On resume, section \ref{Con} provides some concluding remarks and outlines some perspectives. Finally, in the appendix section, we explain frame-independence, polyconvexity, and thermodynamic consistency.
 
\section{Non-linear Features in Cross-Linked Polymers}
\label{CM}

\begin{figure}[H]
\centerline{\includegraphics[width=.95\textwidth]{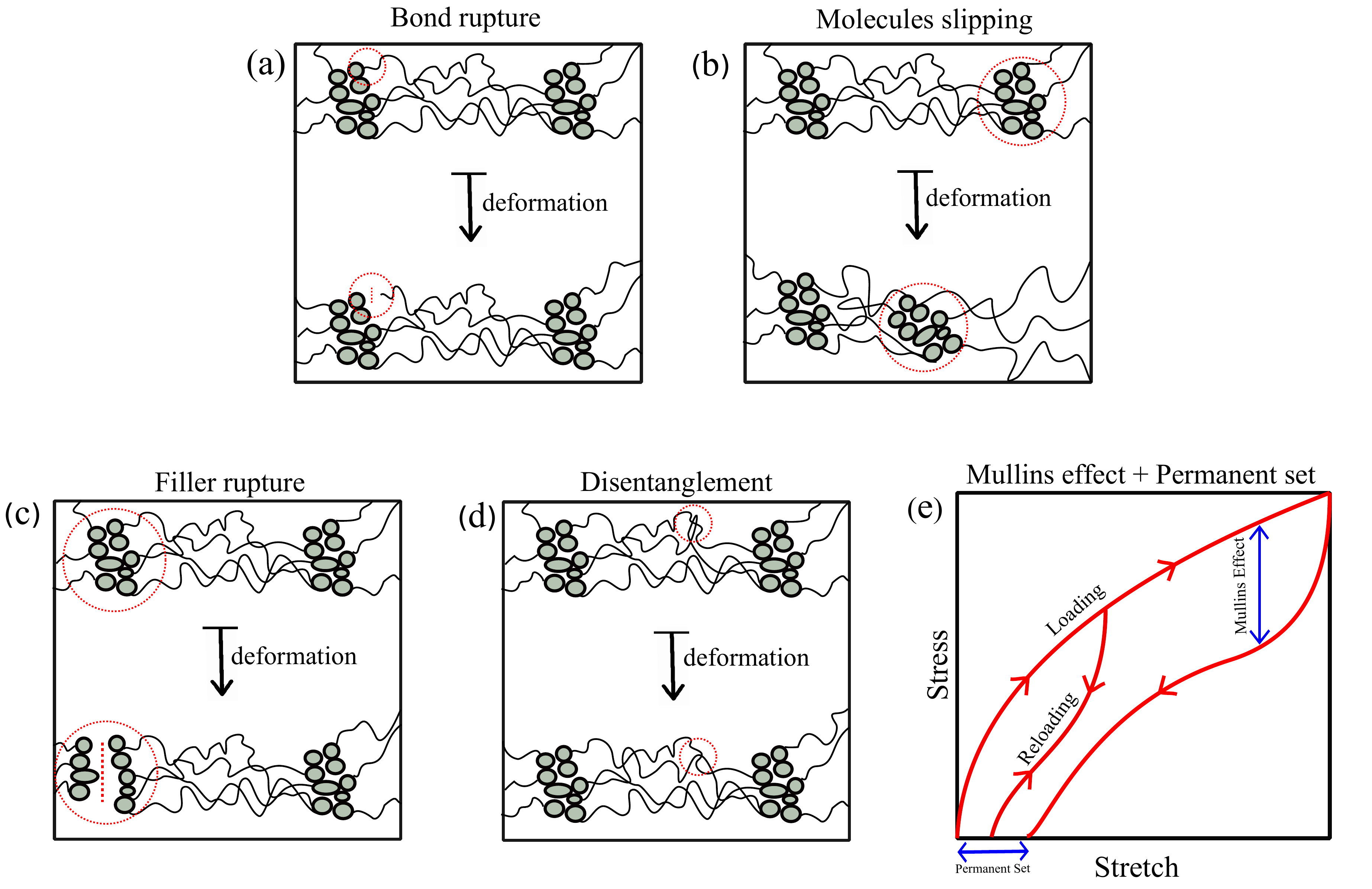}}
\caption{Schematic Physical explanation of Deformation Induced Damage}
\label{Damage}
\end{figure}

Hyper-elasticity in most materials (i.e. cross-linked polymers) is defined by the nonlinear elastic behaviour in  large deformations. 
Meanwhile, cross-linked polymers often exhibit inelastic features in their hyper-elastic behaviour, e.g.  damage after first elongation known as Mullins effect \cite{mullins1969softening, bahrololoumi2020multi}. This phenomenon happens to both types of filled and non-filled cross-linked polymers. To provide a better understanding of micro-structural sources of such inelastic effects, Fig. \ref{Damage} shows some of the  physical sources in amorphous polymeric systems such as chain breakage \cite{bueche1960molecular}, chain disentanglement \cite{hanson2005stress}, molecules slipping \cite{houwink1956slipping} and rupture in cluster of fillers \cite{kraus1966stress}. 
After primary loading, deformation-induced damage often leads to  a residual strain known as permanent set. While the permanent set in unfilled rubber is negligible, it becomes prominent in most filled compounds. Fig. \ref{Damage}.e provides a schematic view on different inelastic features in the hyper-elastic behaviour of a polymeric system.  

To model the inelastic features in the behavior of cross-linked polymers, previous efforts were mostly focused on phenomenological and physics-based approaches.
Physics-based approaches are often excessively complicated for real-time applications, and phenomenological models are not reliable outside design condition. Here, by coupling a physics-based approach to machine learned L-agents, we devise a knowledge-driven ML approach to address inelastic features in hyper-elastic behaviour of cross-linked polymers. {Therefore, to model the nonlinear behavior of cross-linked polymers, we select an appropriate neural network consists of proper activation functions and the number of layers and neurons. Also, to model damages, internal parameters of L-agents are designed based on the type of} materials’ memory. {In material with full memory such as rubbery materials with damage, only the maximum status of history affects the next sequence. Using a micro-sphere as a directional model of polymer matrix guarantees modeling of inelastic features such as permanent set. These steps are explained in the next section in detail.}

\section{Physics-based Reduction}
\label{PDR}

To model the second order stress-strain fields required for characterisation of hyper-elastic material, current approaches ranging from phenomenological  to data-driven, face one major challenge, namely lack of data on 3D structures. There are no tools to measure stress field across a structure, and for strain we can 
only measure the strain field for relatively simple structures using digital image correlation (DIC) techniques \cite{dalemat2019measuring}. 

{Helmholtz free energy $\Psi$ is {a function} of both deformation and temperature. Differentiating $\Psi$ with respect to kinematic state variables at constant temperature yields internal forces (i.e. stresses) defined per unit mass. For the case in which $\Psi$ is solely a function of deformation (i.e. isothermal processes), the Helmholtz free energy referred to as the strain energy function.}

\begin{figure}[H]
\centering
{\includegraphics[width=.9\textwidth]{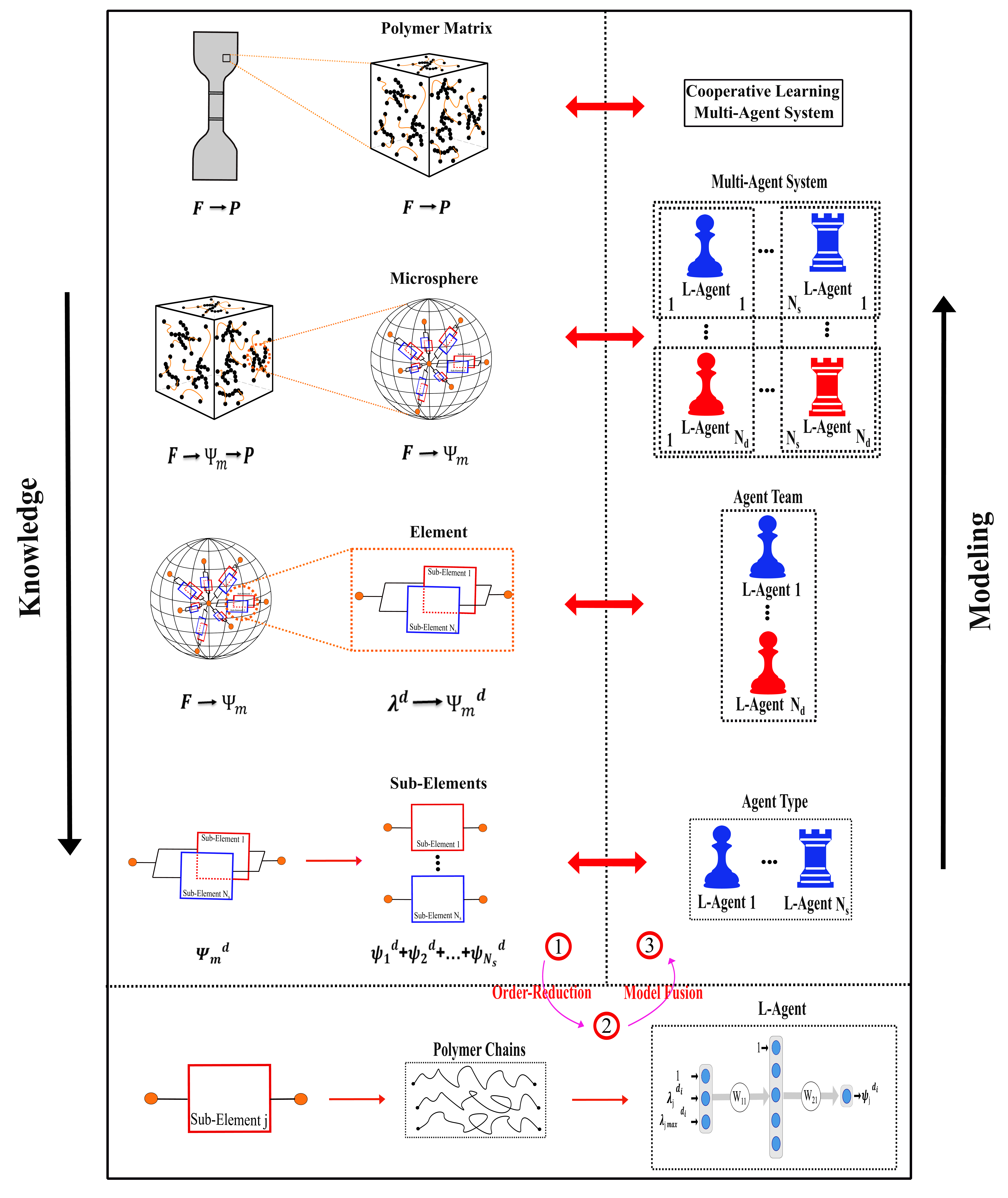}}
\caption{Schematic of proposed model from order-reduction to model fusion}
\label{lit}
\end{figure}

For hyper-elastic materials, strain energy is derived directly from Clausius-Planck form of second law of thermodynamics through different work conjugate pairs, such as  two-point strain/stress tensors ($\tens F$/deformation gradient:$\tens P$/first-order Piola stress), material  strain/stress  tensors ($\tens E$/Lagrange strain:$\tens S$/second-order Piola stress), and spatial  strain/stress  tensors ($\tens L$/Hencky strain:$\boldsymbol{\tau}$/Kirchhoff stress).
Strain energy function must accompany conditions like normalization, growth conditions, isotropy, objectivity, and polyconvexity, which guarantees the uniqueness of the solution (Appendix). In view of the lack of data on the stress fields, a proper modeling approach is expected to be able to only use the limited information obtained from the classical characterization tests on the collective sample behavior. In rare cases, a model can be provided by digital image correlation reconstruction of 2-D strain fields, which shall be used for model validation but should not become a necessary data for model fitting in view of the cost/complexity of the experiment. 
 The challenge of significant missing data has been historically addressed by implementing knowledge of the material behaviour in the model and to constrain the model in advance before having the data. Such a solution is not relevant in data-driven approaches due to lack of infused  knowledge of the material. Here, we propose to address the challenge of significant missing data in high-dimensional data-driven approaches through a physics-driven order-reduction approach by infusing knowledge through implementation of the concept of micro-sphere, network decomposition, continuum mechanics, and polymer physics. Accordingly, we developed a sufficiently constraint machine-learned model that can predict the material behavior solely based on the macro-scale collective behavior of the sample. Fig. \ref{lit} demonstrates a schematic of the proposed model simplification idea.

\subsection{Continuum Mechanics}     
     We introduce the first and the most important constraint from continuum mechanics understanding on 3D mapping of second order stress/strain tensors. While, $\tens F \rightarrow \tens P$ mapping generally needs a  complicated fourth order tensor $\tenf C=\frac{\tens P}{\tens F}$, in hyper-elastic materials, unlike hypo-elastic materials, the procedure can be simplified using an internal energy density function. Accordingly, we can use Finite strain theory to simplify $\tens F \rightarrow \tens P$  mapping by introduction of the strain energy $\Psi_m$  as the middle agent in mapping, where $\tens F \rightarrow \Psi_m  \rightarrow  \tens P$. The strain energy is a non-negative scalar-valued function $\Psi_m(\tens F)$ which can replace part of the process requires to derive tensor-valued stress function $\tens P(\tens F)$. The increment of $\Psi_m$ denotes the stress required to change the strain field, and thus $\Psi_m$ can be described with respect to any stress-strain work conjugates such as (i) two-point tensors, (ii) material tensors, or (iii) spatial tensors, as shown below
      \begin{equation}
\label{conjugate}
    \tens{P} = \frac{\partial \Psi_m}{\partial \tens F}, \qquad \tens{S}=\frac{\partial \Psi_m}{\partial \tens E}, \qquad  \boldsymbol{\tau}=\frac{\partial \Psi_m}{\partial \tens L}.
\end{equation}
      One particular advantage of using $\Psi_m$ as middle agent is that it ensures the material objectivity, and thermodynamic consistency on all the derived constitutive model (see Truesdell et al. \cite{truesdell2012elements}). Considering the physics of the problem, certain restrictions exist for strain energy which needs to be enforced further in the data-driven model, namely
            \begin{align}
            \label{const1}
   \Psi_m(\tens F) \geq 0 \quad &\textit{when} \quad  \tens F \ne 0 \qquad \qquad\quad\textit{Increase energy by deforming},          \nonumber\\
\Psi_m(\tens F)=0 \quad &\textit{when} \quad  \tens F=\tens I \qquad \quad \qquad \textit{Normalization condition},  \\
\Psi_m(\tens F) \rightarrow \infty \quad &\textit{when} \quad det \tens F\rightarrow \infty/0 \qquad \textit{Growth condition} \nonumber.
\end{align}

      Further restrictions can be introduced by finite strain theory to ensure stability of $\Psi_m$ in large deformations of certain materials. For hyper-elastic materials, ellipticity is a major concern which can be enforced  by verifying the strain energy in the absence of traction forces in two arbitrary directions \cite{holzapfel2000new}. Verifying this condition is generally labor-intensive, so polyconvexity is introduced as a stronger condition that entails ellipticity. It is also simpler to verify \cite{hartmann2003polyconvexity}, as discussed in the appendix.     \textit{So, the first constraint that we enforce in our model, is enforcing agents to derive $\Psi_m(\tens F) $ such that it satisfies Eq. \ref{const1} and polyconvexity condition.
      }
      
\subsection{Micro-Sphere} 

\begin{figure}[H]
\centering
{\includegraphics[width=.3\textwidth]{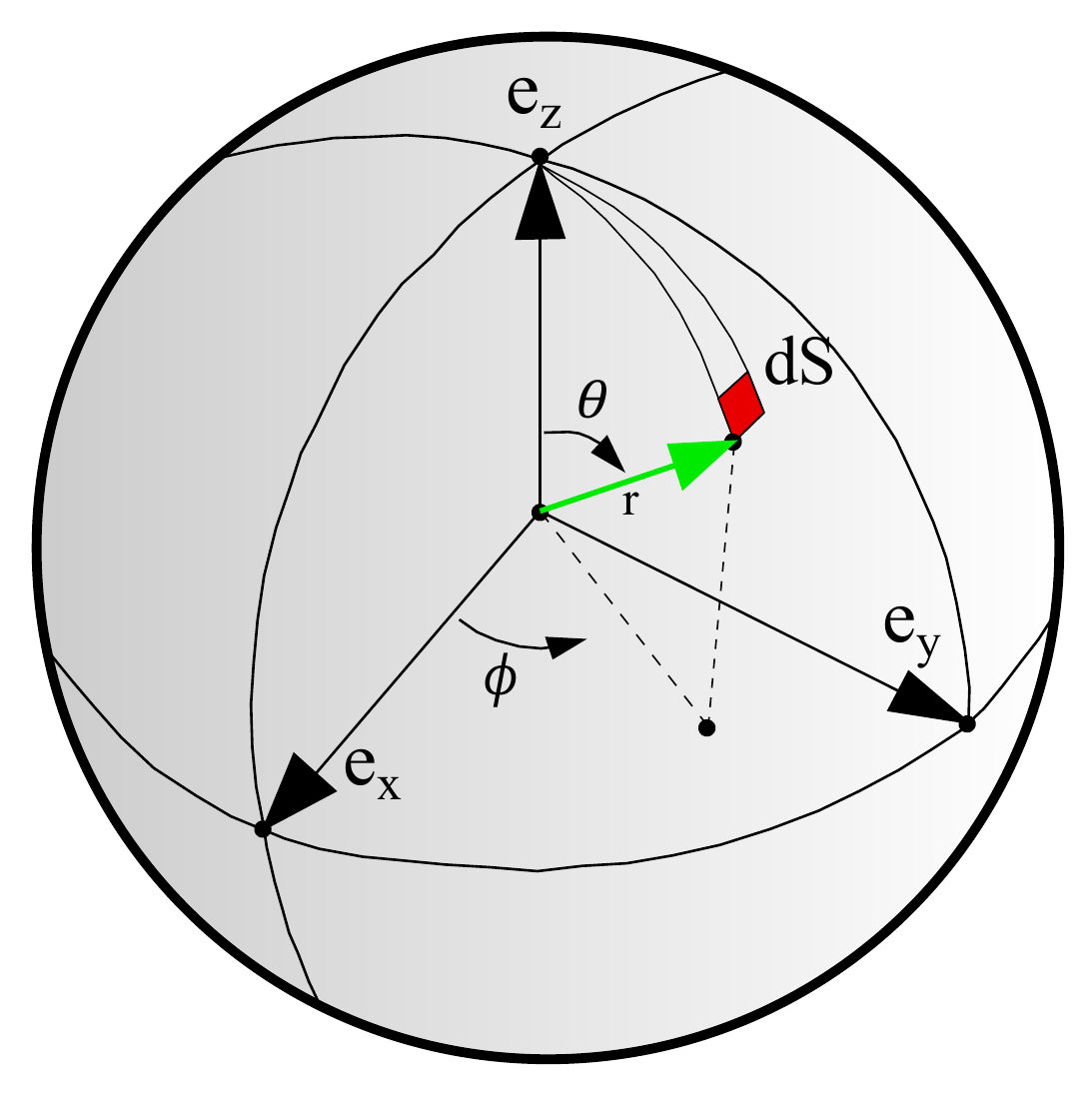}}
\caption{The unit micro-sphere and the orientation vector in terms of spherical coordinates}
\label{microspherefig}
\end{figure}

      {The} second constraint is implemented based on polymer physics, in particular topology of cross-linked amorphous network. Knowing amorphous systems are isotropic at virgin state, polymer chains are considered to be uniformly distributed in all spatial directions. Such homogenized spatial arrangement of polymer chains allow us to use  the micro-sphere concept to represent the 3D matrix as a homogeneous assembly of similar 1D elements that are distributed in different spatial directions over a micro-sphere (see Fig.~\ref{lit}).  This approach can transfer information from super-simplified 1D elements to generate complex 3D behavior of the matrix via homogenization over the unit-sphere. Furthermore by discretizing the sphere into  finite sections, the integration can be taken out numerically over $N_d$ integration directions ${\left[{\vect{d}_{i}}\right]_{i=1...N_d}}$ with different weight factors ${[w_{i}]}_{i=1...n}$ \cite{bavzant1986efficient}. Accordingly, strain energy of the matrix $\Psi_m $ with respect to its elements can be written as

\begin{equation}
\label{direct}
    \Psi_m   ={\frac{1}{4\pi}} \int_S {\mathop {\Psi_{m }^{ \vect{d}}}} dS^{^{ \vect{d}}} \cong {\sum_{i =1}^{N_d} {w_i} {\mathop {{\Psi}_{m }}}^{ \vect{d_{i}}} }, \qquad  \textit{where} \qquad  
     {\mathop {{\Psi}_{m }}}^{ \vect{d_{i}}}=   \mathcal B^{ {\vect{d}_{i}}}
\end{equation}
where ${\mathop{{\Psi}_{m}}}^{ \vect{d_{i}}} $ is the energy of sub-matrix element in direction $ {\vect{d}_{i}}$ which will be represented by one team of L-agent  $\mathcal B^{ {\vect{d}_{i}}}$ which represents an additive cooperation between multiple L-agents $\mathcal A_\bullet ^i $. Eq\ref{direct} represents the integral $S(\theta, \phi)= \int_{0}^\theta \int_{0}^\phi \sin{(\theta)} d\theta d\phi$ over the unit-sphere with the unit vector $\vect{r}=\sin{(\theta)} \cos{(\phi)} \vect{e}_x + \sin{(\theta)} \sin{(\phi)} \vect{e}_y + \cos{(\phi)} \vect{e}_z$ (see Fig. \ref{microspherefig}). Assuming identical team in all directions in the virgin state, namely  $\mathcal B^{ {\vect{d}_{i}}}=\mathcal B^{ {\vect{d}_{j}}}$, initial isotropy is assured, although the material can quickly become anisotropic due to different loading on different directions. Moreover, since L-agents react to varying loading in each direction, the model can consider the onset of damage, deterioration, and propagation of cascading failure in materials with directional response.

\subsection{Network Decomposition}      
      {The} third constraint is derived from statistical mechanics, namely by infusing the concept of superposition, which allow us to predict complicated patterns by superposing simple patterns on top of each other. The concept, {a.k.a} network decomposition concept in constitutive modeling \cite{dargazany2014generalized}, will be carried out by representing the energy of an element,${\mathop{{\Psi}_{m}}}^{ \vect{d_{i}}} $ by superposing the energy of multiple sub-elements,  ${\mathop{{\Psi}_{m}}}^{ \vect{d_{i}}} = \sum_{j=1}^{N_s}{\mathop{{\Psi}_{j}}}^{ \vect{d_{i}}} $, where each sub-element is responsible for one simple inelastic feature. Representing each sub-elements by one L-agent, we can calculate the energy of one element by a team of cooperative L-agents $\mathcal B^{ {\vect{d}_{i}}}=[ \mathcal A_j^i  ]$, and then replicating this cooperative team in different directions to provide us with the energy of the matrix. To this end, by substituting  Eq.\eqref {direct}, we can directly derive the energy of the matrix with respect to sub-elements and the L-agents which represent them as given here
      
\begin{align}
\label{direct1}
    \Psi_m  &={\frac{1}{4\pi}} \int_S {\mathop {{{\Psi_m}^{ \vect{d}}}}} dS^{^{ \vect{d}}} \cong 
\sum_{i=1}^{N_d} \sum_{j=1}^{N_s} {w_{i}} {\mathop{{\psi_j}}}^{ \vect{d_{i}}} \nonumber \\
 \Psi_m  &\approx \sum_{i=1}^{N_d} \sum_{j=1}^{N_s} {w_{i}}  \mathcal A_j ^i \qquad  \textit{where} \qquad  
     {\mathop {{\Psi}_{m }}}^{ \vect{d_{i}}}= \sum_{j=1}^{N_s}  \mathcal A_j ^i.
\end{align}
where $N_s$ is the number of sub-elements considered for each element. Consequently, we derived super-simplified scalar-to-scalar mapping behaviour for each element which can be represented simplified 2-layer feed-forward neural network L-agent $\mathcal A_j ^i $. 
While training data are only available on collective behaviour of the L-agents, the input parameters can be defined for each L-agent  team individually.
Each L-agent, $\mathcal A_j^i:={\psi_j^i}(\tenf{E}^i, \tenf{M}_j)$, will be trained based on a set of non-kinematic input $\tenf E^i$ and internal $\tenf M_j$ parameters, which depending on material memory (full or recent), can satisfy normalization, growth conditions, isotropy, objectivity, and polyconvexity. 

The input vector is independent of the sub-element definition and should represent the problem setting, material or loading, e.g. stretch \& time. Internal parameters are specifically hypothesized for the model to capture the evolution of damage and vary for each network. The behavior of all teams should be identical in the virgin state to represent initial isotropy, so one has $\mathcal A_j ^i =  \mathcal A_j ^k  \forall i\neq k $. Accordingly, we only assume different ANN types for L-agents associated to different sub-elements. All replicated  agents associated to one  sub-element are the same despite being distributed in different teams to represent different directions. For the replicated agents,  only the  inputs are different depending on their direction (see Fig.\ref{lit}).So the energy of one sub-element can be written as
\begin{equation}
{ \psi_{j}^{ \vect{d}_i}} =  \mathcal A_j ^i= {{ANN}_{j}} (\tens{W_{j}},\tenf{E}^i, \tenf{M}_j), 
\end{equation}
where $\vect{W}_{j}$ is the weight vector associated to L-agent   $\mathcal A_j ^\bullet$, and $\tens{W}=\left[ \vect{W}_{1}... \vect{W}_{N_s}\right]$ is the weight matrix representing assembly of all  $\vect{W}_{j}$. Consequently, based on Eqs. \ref{conjugate} and \ref{direct1}, the first Piola-Kirchhoff stress tensor $\tens P$ can be derived as

\begin{equation}
\label{stress}
\tens{P}= \frac{\partial{\Psi_m}}{\partial {\tens{F}}}-{p}{{\tens{F}}^{-T}}
\tens{P}=  \sum_{i=1}^{N_d} \sum_{j=1}^{N_s} {w_{i}} \frac{\partial{ {\mathcal A_j ^ i}}}{\partial {\tens{F}}}-{p}{{\tens{F}}^{-T}}, 
\end{equation}
where $p$ denotes the Lagrange multiplier to guarantee incompressibility of the material. To train the model, a cost function should be derived to quantify prediction error against experimental observations on collective sample behaviour, e.g.   uni-axial tensile test  provides   1D dataset $\mathcal S=[\tilde P, \tilde \lambda]$, with nominal stress $\tilde P$ and stretch $\tilde \lambda$ in direction of principal stretch. Here, the error has  been  quantified using  least-square method by writing

\begin{equation}
\label{error}
\small{
E(\tens {W}) =\frac{1}{2} {\sum_{n=1} \left[ \vect{g}_1 (\sum_{i=1}^{N_d} \sum_{j=1}^{N_s} {w_{i}}\frac{\partial{{{\psi_j}}^{ \vect{d_{i}}}}}{\partial {\tens{F}}}-{p}{{\tens{F}}^{-T}})\vect{g}_1
-P_n\right]^2}},
\end{equation}
which $P_{(1,1)}:=\vect{g}_1\tens P\vect{g}_1$ is the first component of  the experimental macro-scale stress tensor $\tens P$ in loading direction $\vect{g}_1$.

\textbf{Neural Network L-agents:} 
Artificial neurons,  also known as nodes, are the basic units in the neural system, which receive and transfer information to the other nodes through activation functions. Accordingly, the ability of "learning" of an ANN agent strongly depends on proper selection of the activation function for each node. Their purpose is to decide whether a neuron should be activated or not and introduce non-linearity into the output of a node. Therefore, they make the model to generalize or adapt with a variety of data and to differentiate between the output. The cost function is accounted for estimating $\tens {W}$ using gradient descent (GD) algorithm, which gradually optimize the initial guess toward target values.

\textbf{Material With Full or Recent Memory:} For history-dependent materials,  parameters should be specifically chosen to represent material memory and then fed into the L-agents through internal parameters. However, different type of memory parameters may be required for describing different materials, e.g. for materials with recent memory such as visco-elastic materials, internal parameters should transfer information from each iteration to the next. In contrast, for material with full memory such as elastomers, the internal parameters can be defined independent of the solution iterations as damage-precursor of the external events, for example the maximum stretch in rubber material can be used as a damage precursor to show the history of maximum loading in each direction.

\section{Implementation to Rubber Inelasticity}
\label{IRI}
To show the performance of the proposed model, inelastic behavior of rubber has been studied. The  number of teams and their associated agents can be chosen based on the trade-off between accuracy and computational cost. Here,  we choose 21 teams, each with two agents which is a relatively small number  \cite{bavzant1986efficient} (21 integration-point discussed  in Appendix).  
The inputs and internal parameters of L-agents are designed to capture the rubbers deformation with full memory through $\lambda_{j-max}$ parameters.
To enable teams to predict different states of deformation, each teams should be provided with the  first and second invariants of deformation\cite{lambert1999new}. The condition is satisfied by providing {input set $ \vect S_1^{d_i} = [ \lambda_1^{\vect{d_{i}}} ; {\lambda_{1-max}^{\vect{d_{i}}}}]$   to L-agent 1 and $ \vect S_2^{d_i} = [\lambda_2^{\vect{d_{i}}} ; {\lambda_{2-max}^{\vect{d_{i}}}}]$ to L-agent 2}
\begin{equation}
    \label{lambda}
{\lambda_1}^{\vect{d_{i}}}=\sqrt{{\vect{d_{i}}}\tens{C}{\vect{d_{i}}}},  \qquad 
{\lambda_2}^{\vect{d_{i}}}=\sqrt{{\vect{d_{i}}}{\tens{C}}^{-1}{\vect{d_{i}}}},  \qquad 
\tens{C}={\tens{F}}^{T}{\tens{F}}
\end{equation} 
where ${\lambda_1}^{\vect{d_{i}}}$ and ${\lambda_2}^{\vect{d_{i}}}$ are designed that lead to first and second sub-elements represent the $I_1$ and $I_2$, respectively.
For the ANN structure of L-agents, we consider one input layer, one hidden layer with four neurons and three activation functions soft plus ($\psi(\bullet)=ln(1+e^{\bullet})$), sinusoid ($\psi(\bullet)=sin(\bullet)$) and hyperbolic tangent ($\psi(\bullet)=tanh(\bullet)$). In summary, we represented the rubber matrix by the cooperative game of 21 teams of 2 agents through  $\mathcal{A}^i_j, \,\, i \in \left\{1,21\right\},\,\, j\in \left\{1,2\right\}$.  Final cost function after agents fusion is given by 

\begin{equation}
\small
\label{error1}
E(\tens W_1, \tens W_{2}) =\frac{1}{2} {\sum_{n=1}} [\vect{g}_1({\sum_{i = 1}^{21} {\sum_{j = 1}^{2}} {w_i} {\frac{\partial \mathcal{A}^i_j}{\partial {\lambda_j}^{ \vect{d_{i}}}}}  {{\frac{\partial {{\lambda_j}^{ \vect{d_{i}}}}}{\partial {\tens{F}}}}}}-{p}{{\tens{F}}^{-T}})\vect{g}_1 
-P_n]^2,  
\end{equation}
subjected to weights related to ${{\lambda_{1-max}}}$  and  ${{\lambda_{2-max}}}$  $\leqslant 0$ ; and weights related to ${\lambda_1}$  and  ${\lambda_2 }$   $\geqslant 0$ to satisfy thermodynamic consistency and polyconvexity respectively. Eq. \ref{derive1} and Eq. \ref{derive2} show the derivation of each sub-element's energy with respect to deformation gradient. Accordingly, Fig. \ref{lit} shows the schematic concept of the derived model.

\begin{equation}
    \label{derive1}
     {\sum_{i = 1}^{21} {w_i} {\frac{\partial \mathcal{A}^i_1}{\partial {\lambda_1}^{ \vect{d_{i}}}}}  {{\frac{\partial {{\lambda_1}^{ \vect{d_{i}}}}}{\partial {\tens{F}}}}}}  =  {\sum_{i = 1}^{21} {w_i} {\frac{\partial \mathcal{A}^i_1}{\partial {\lambda_1}^{ \vect{d_{i}}}}} {\frac{1}{{\lambda_1}^{ \vect{d_{i}}}}} {\tens{F}} \left({{ \vect{d_{i}}} \otimes { \vect{d_{i}}}}\right)}.
\end{equation}

\begin{equation}
    \label{derive2}
     {\sum_{i = 1}^{21} {w_i} {\frac{\partial \mathcal{A}^i_2}{\partial {\lambda_2}^{ \vect{d_{i}}}}}  {{\frac{\partial {{\lambda_2}^{ \vect{d_{i}}}}}{\partial {\tens{F}}}}}}  = - {\sum_{i = 1}^{21} {w_i} {\frac{\partial \mathcal{A}^i_2}{\partial {\lambda_2}^{ \vect{d_{i}}}}} {\frac{1}{{\lambda_2}^{ \vect{d_{i}}}}} {{\tens{F}}^{-1}} {{\tens{F}}^{-T}} {{\tens{F}}^{-1}} \left({{ \vect{d_{i}}} \otimes { \vect{d_{i}}}}\right)}.
\end{equation}

\subsection{Minimizing Data Requirement for Training}
\paragraph{Data-set Minimization}
{A critical step in the selection of the training dataset is to understand the role of the training points and assure their quality in the model predictions.} Too little data may provide a false sense of confidence by preventing us to see the critical points, while low-quality data may provide faulty results which seems perfectly robust.
For example, in the aforementioned  model  developed for rubber, we have introduced two L-agent types which represents two sub-elements using $\vect S_1^{d_i}$ and $\vect S_2^{d_i}$ input sets, respectively. In view of the definition of $\vect S_2^{d_i}$, we know that it has a limited variation in uni-axial tensile loading which makes the contribution of the second L-agent almost limited in such loading. However, $\vect S_2^{d_i}$ significantly varies in bi-axial loading, which makes the contribution of second L-agent quite considerable in this case.
Therefore, training with uniaxial data cannot provide quality information needed for confident training of both agents since second-agent cannot be fully engaged.
 
 \textit{In essence, we cannot train agents with the scenarios that they are not participating in or have minimal contribution. Thus, the confidence in  training of agents is directly correlated to the quality of the training data, and in contribution of agents in those scenarios.} However, by defining the quality of data with respect to the input required by each agent,  we can quantify the \textbf{confidence interval} in which an agent can be trained with high confidence  with respect to the provided data. 
 
{The} confidence interval of a system is equal to that of its  agent with least confidence. {The} confidence of an agent can be calculated with respect to the deformation range used  in each direction for training of that agent.   Since the reliability of the predictions of each agent is related  to  its training, we can linearly correlate the agent's reliability to their training range. As an example, in case of uniaxial tension where the sample is stretched till $\chi_{x}$, maximum first deformation state (axial) is $\chi_{x}$ which occurs in the loading direction, and minimum is $\frac{1}{\sqrt{\chi_{x}}}$, which occurs in the transverse directions. Similarly, the training domain for the second agent is $[\frac{1}{\chi_x},\sqrt{\chi_x}]$. 
In case of bi-axial tension, range of agents deformations are $[\frac{1}{\chi_{bi.}^2},\chi_{bi.}]$ and $[\frac{1}{\chi_{bi.}},\chi_{bi.}^2]$. 

\begin{figure}
\center{\includegraphics[width=0.9\textwidth]{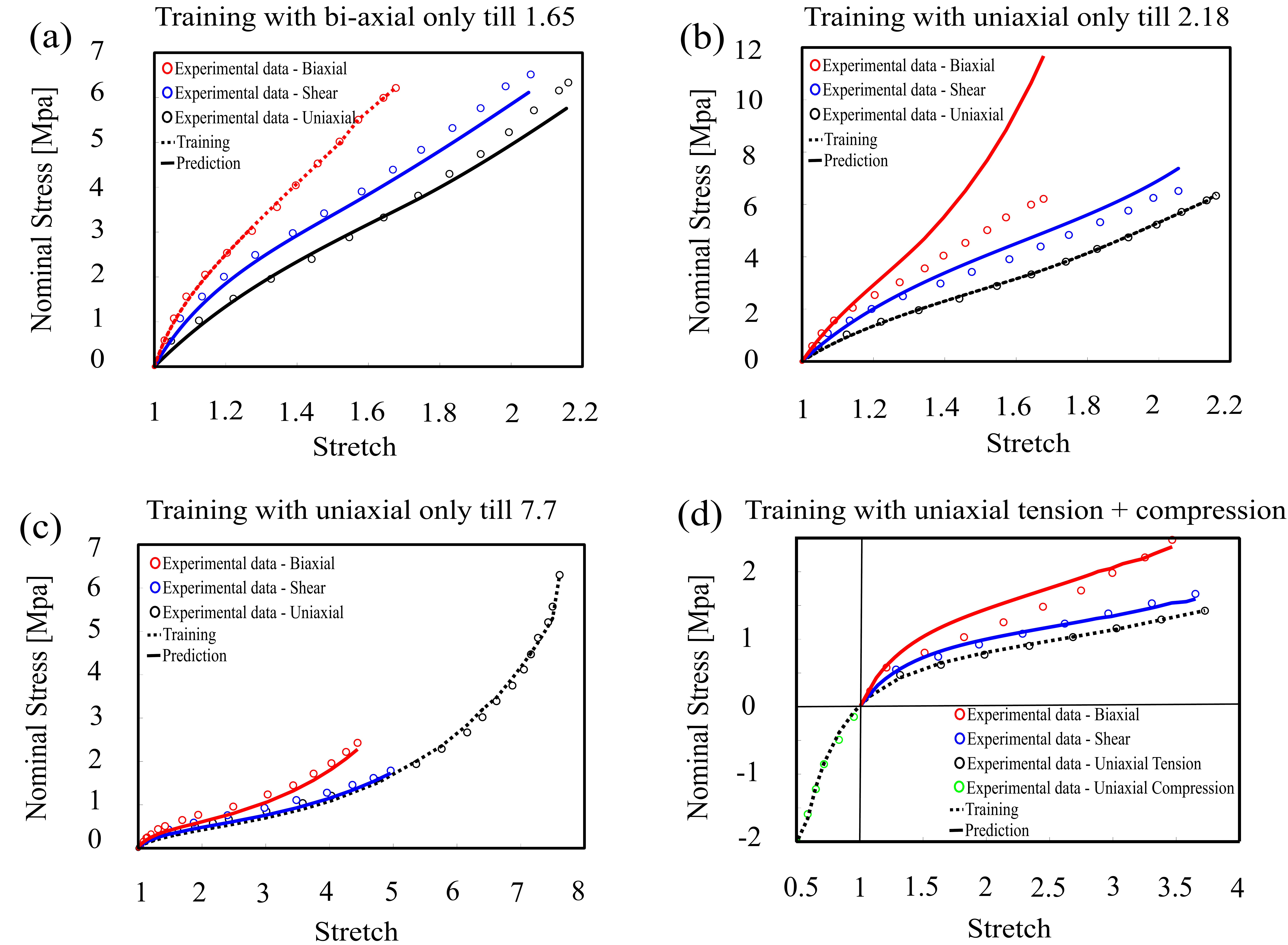}}
\caption{Model training and prediction with a) bi-axial tension training (filled natural rubber) b)uniaxial tension training (data set\cite{mars2004observations}) c) uniaxial tension training (Treloar's data set \cite{treloar1975physics}) d) uniaxial tension and compression training (data set \cite{heuillet1997modelisation})}
\label{sphere2}
\end{figure}

\begin{figure}
\centerline{\includegraphics[width=0.9\textwidth]{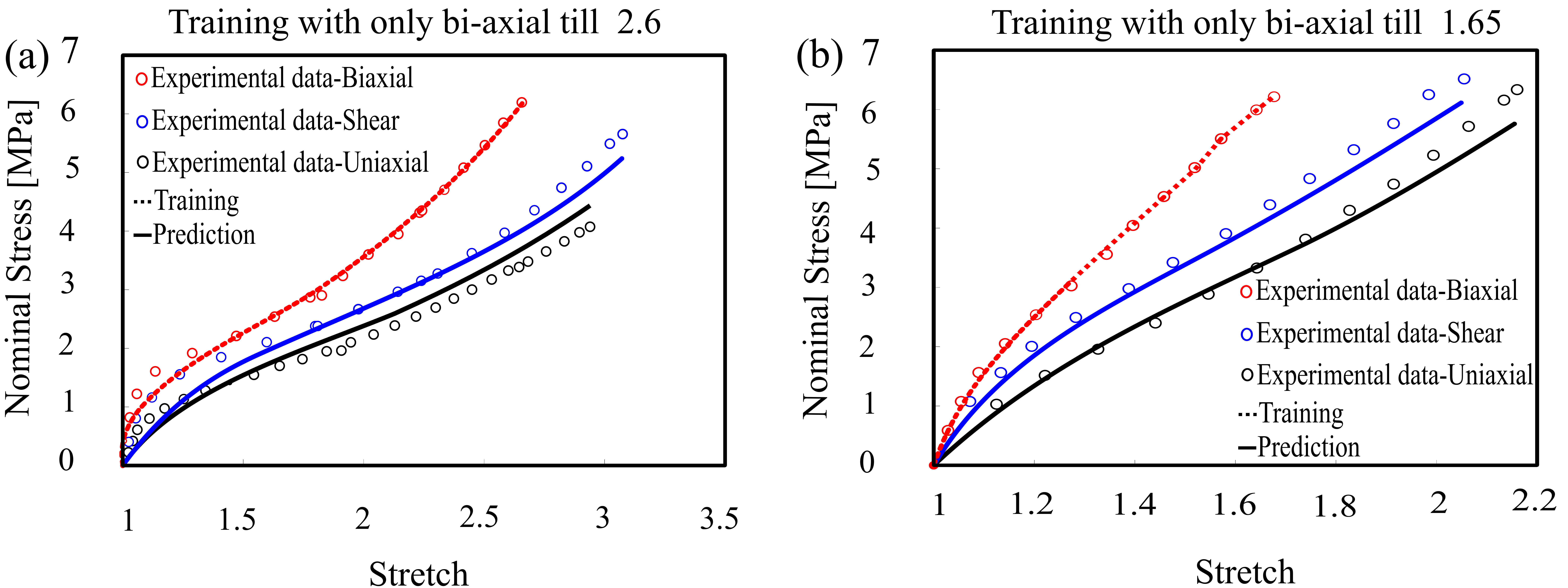}}
\caption{Model training and prediction of uniaxial, bi-axial and pure shear a) Urayama's data set \cite{mai2017novel} b) Mars's data set \cite{mars2004observations}}
\label{loading}
\end{figure}

If we train the model based on uniaxial tensile data till $\chi_{x}$, the model can predict different states of deformation based the ranges that the model have calibrated based on that. In order to ensure accurate prediction of the model, the prediction ranges should be in the range that agent is trained. Thus, in the bi-axial prediction case, the model limits to
\begin{align}
    \textit{Agent 1:} \quad \quad \quad \left[\frac{1}{\chi_{bi.}^2},\chi_{bi.}\right] \in \left[\frac{1}{\sqrt{\chi_{x}}},\chi_{x}\right] \longrightarrow \chi_{bi.}\le \sqrt[4]{\chi_{x}}\nonumber
    \\
    \textit{Agent 2:} \quad \quad \quad \left[\frac{1}{\chi_{bi.}},\chi_{bi.}^2\right] \in \left[\frac{1}{\chi_x},\sqrt{\chi_x}\right] \longrightarrow \chi_{bi.} \le \sqrt[4]{\chi_{x}}.\label{Conf_int}
\end{align}
{As} it can be in Eq. \ref{Conf_int}, these two ranges result into a same confidence interval for the agents. Accordingly, considering one of these agents confidence interval would be sufficient to calculate the network reliability. 
 Likewise, these training/prediction domains can be calculated for different cases of traing and predictions with different states of deformation, see table \ref{tab}. To show the confidence interval, we explore five different training data set and their resulting agents. Using two set of experiments for the training purpose can increase the predictability range of the model as each of the experiments can be activated in different ranges and agents. Note that the model may extrapolate and predict more than confidence interval but it is not necessarily accurate. 
 \begin{table}[t]
\centering 
\small
\caption{{Prediction domain for train till stretch $\chi$}}
\label{tab}
\begin{tabular}{l|*{6}{c}}\hline 
\backslashbox{Prediction}{Training}
&\makebox{Uni. \footnote{Uniaxial} Tensile}&\makebox{Bi. \footnote{Biaxial}}&\makebox{Pure Shear}&\makebox{Uni. Comp. \footnote{compression}}&\makebox{Plane Strain Comp.}\\\hline 
\textbf{} \quad \quad  Uni. Tensile  & $\chi$ & $\chi$ & $\chi$& $\frac{1}{\sqrt{\chi}}$ & $\frac{1}{\chi}$\\\hline
\textbf{} \quad \quad \quad \quad  Bi. & $\sqrt[4]{\chi}$ & $\chi$ & $\sqrt[4]{\chi}$& $\frac{1}{\sqrt{\chi}}$ & $\frac{1}{\chi}$  \\\hline
\quad \quad  Pure Shear & $\chi$ & $\chi$ & $\chi$& $\frac{1}{\sqrt{\chi}}$ & $\frac{1}{\chi}$\\\hline
\quad  \qquad Uni. Comp. & $\frac{1}{\sqrt{\chi}}$ & $\frac{1}{\chi^2}$ & $\frac{1}{\chi}$& $\chi$ & $\chi$\\\hline
\quad  Plane Strain Comp. & $\frac{1}{\sqrt{\chi}}$ & $\frac{1}{\chi^2}$ & $\frac{1}{\chi}$& $\chi$ & $\chi$\\\hline
\end{tabular}
\end{table}
\begin{enumerate}
	\item \textbf{ Training with uniaxial only vs biaxial only} Mars dataset which has three modes of pure shear, uni-axial and bi-axial tensile tests have been used  \cite{mars2004observations}. In first case, the model is trained by  bi-axial data only till $\chi=1.65$ and validated against other modes (see Fig. \ref{sphere2}.a). Confidence interval in uniaxial and pure shear predictions is also limited to $\chi=1.65$.  
	
	In second case, the model is trained by  uniaxial data only till $\chi=2.18$ and validated against other modes (see Fig. \ref{sphere2}.b). Confidence interval in shear will be limited to  $\chi=2.18$  but in bi-axial will be dramatically reduced to $\chi=1.21$ due to the uncertainty in training L-agent 2.

		\item \textbf{ Training with uniaxial only over a long range}  Here, we showed that we can improve the confidence interval  of one agent not only by choosing the games in which it has high contribution, but also by increasing the length of the game in which one agent has small contribution. \textit{In essence,  we can have a short game with high contribution, or long game with low contribution.} In case of rubber, uniaxial tension is a game in which 2nd L-agent has low contribution. So, here we show that for a sufficiently long game (uni-axial till $\chi=7.7$), we can increase the confidence interval for the second L-agent (bi-axial till $\chi=1.66$), see Fig. \ref{sphere2}.c Treloar dataset which has three modes of pure shear, uni-axial and bi-axial tensile tests have been used  \cite{treloar1975physics}.
	
		\item \textbf{ Training with uni-axial Tension and Compression}  Here, we showed that we can improve the confidence interval by using multiple games to train the agents. So, here model is trained by uniaxial tensile (till $\chi=3.7$) and compression data (till $\chi=0.4$). The confidence in training  of the 1st L-agent is mainly defined by the uni-axial tensile test while that of the 2nd L-agent is formed by compression test.
		The predictions of the trained agents were validated against other modes (see Fig. \ref{sphere2}.d), and as expected confidence interval in bi-axial till $\chi=1.58$ and pure shear predictions is also limited to $\chi=3.7$.  
		Heuillet data-set with three modes of pure shear, uniaxial and bi-axial tensile tests have been used for training/validation  \cite{heuillet1997modelisation}.
	
\end{enumerate}
\paragraph{Accuracy within Confidence Interval}
The proposed engine shows exceptional accuracy within the confidence interval which is comparable to some of the most comprehensive and most expensive knowledge-based models.  We have shown the predictions of the models against different sets of data provided by Uramaya\cite{mai2017novel} and Mars \cite{mars2004observations}, where the model were trained using bi-axial tests only, see Fig. \ref{loading}. Bi-axial tests were chosen to provide the longest confidence interval for other modes (see table \ref{tab}).
 
To show the performance of the proposed model, we compared the relative error of our model in fitting and prediction of Treloar's data set with the non-affine micro-sphere model \cite{miehe2004micro}, WYPiWYG model \cite{amores2019average}, and network averaging tube model \cite{khiem2016analytical}. Note that the error reported for the non-affine micro-sphere model and network averaging tube models \textbf{is fitting error not prediction error}, since they have used all three uniaxial, bi-axial, and pure shear at the same time in their published results. 

Although the proposed model and WYPiWYG model use uniaxial data for training and predict other states of deformation. Thus, results show the excellent performance of our model; however, the proposed model is not complicated and data-dependent as much as other physics-based models (see table \ref{tabcom1}).
\begin{table}[h]
\centering
\small
\caption{{Relative error for several well-known models for Treloar's data set}}
\label{tabcom1}
\begin{tabular}{c c c c c}
\hline
Model Type   & AI             & Phenomenological & \multicolumn{2}{c}{Micro-Mechanical}                                                                                                                  \\ \hline
             & Proposed model & WYPiWYG model \cite{amores2019average} & \begin{tabular}[c]{@{}c@{}}Non-affine\\ micro-sphere model \cite{miehe2004micro}\end{tabular}    & \begin{tabular}[c]{@{}c@{}}Network averaging\\ tube model \cite{khiem2016analytical}\end{tabular}    \\ \hline
Error(\%)    & 1.12           & 5.26             & 0.93                                                                       & 2.11                                                                      \\ \hline
Training Set & Uniaxial       & Uniaxial         & \begin{tabular}[c]{@{}c@{}}Uniaxial + Pure shear\\ + Bi-axial\end{tabular} & \begin{tabular}[c]{@{}c@{}}Uniaxial + Pure shear\\ +Bi-axial\end{tabular} \\ \hline
\end{tabular}
\end{table}

\begin{figure}[H]
\centerline{\includegraphics[width=.45\textwidth]{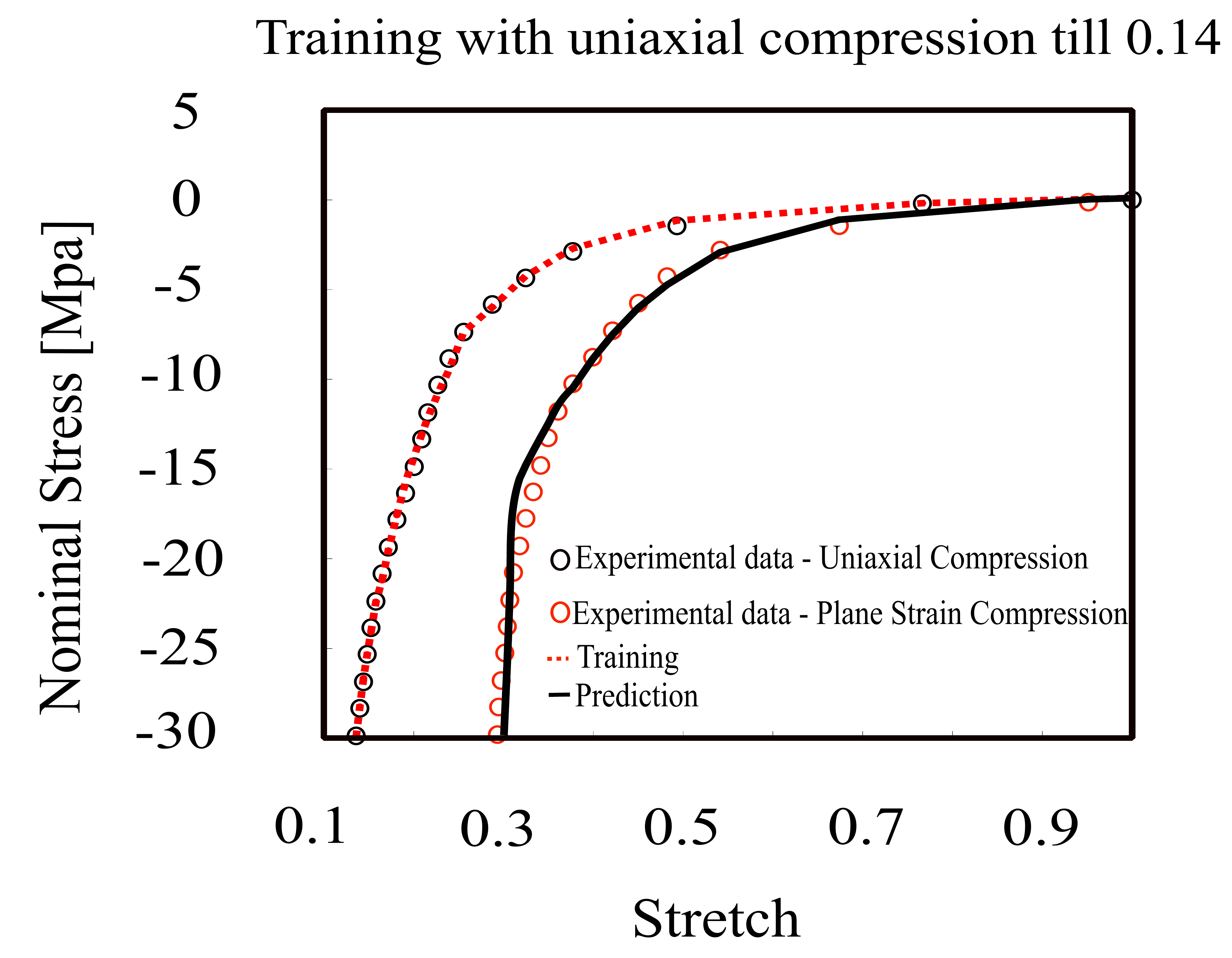}}
\caption{Model training with uniaxial compression and prediction of plane strain compression (Arruda-Boyce's data set \cite{arruda1993three})}
\label{comp2}
\end{figure}

The compression behavior of rubber-like material is another aspect that plays an essential role in industrial application. We trained the model with the data set of uniaxial compression experiments and predicted the behavior of plane strain compression. Fig. \ref{comp2} shows the performance of the proposed model for compression tests. The error in training and prediction of the proposed model for Arruda-Boyce data is $ 0.73 \%$, which compared to the non-affine micro-sphere model, which has $1.29\%$ error, shows a significant performance of our less complicated model.

\begin{figure}[H]
\centerline{\includegraphics[width=1.1\textwidth]{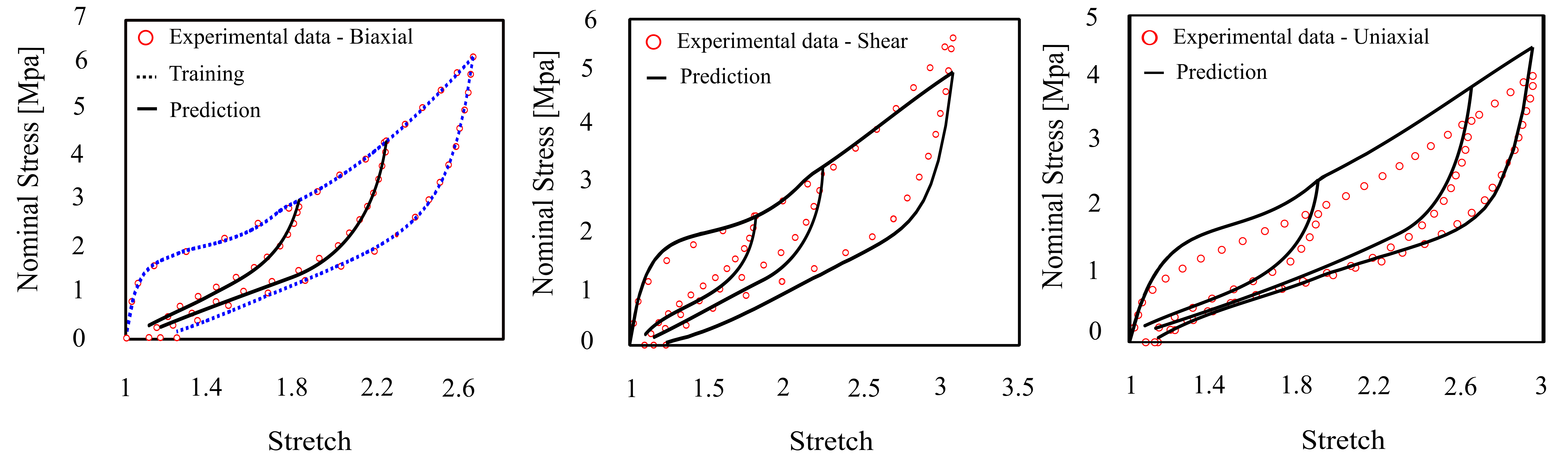}}
\caption{Model training and prediction of uniaxial, bi-axial and pure shear (Urayama's data set \cite{mai2017novel})}
\label{inelastic}
\end{figure}

\paragraph{Damage Prediction and Deformation History}
To further investigate the performance of the proposed model in material with full memory, we predicted  the inelastic features in the behaviour of filled elastomer, namely Mullins effect and permanent set. Fig. \ref{inelastic} shows that stress-stretch curves for this cross-linked polymer with experimental data of \cite{mai2017novel}. We used one set of bi-axial loading-unloading  till $\chi=2.7$ for training and predict inelastic effects in different states of deformation e.g. uniaxial and pure pure shear at increasing  stretch amplitudes which constitutes deformation histories.

\begin{figure}[H]
\centerline{\includegraphics[width=0.9\textwidth]{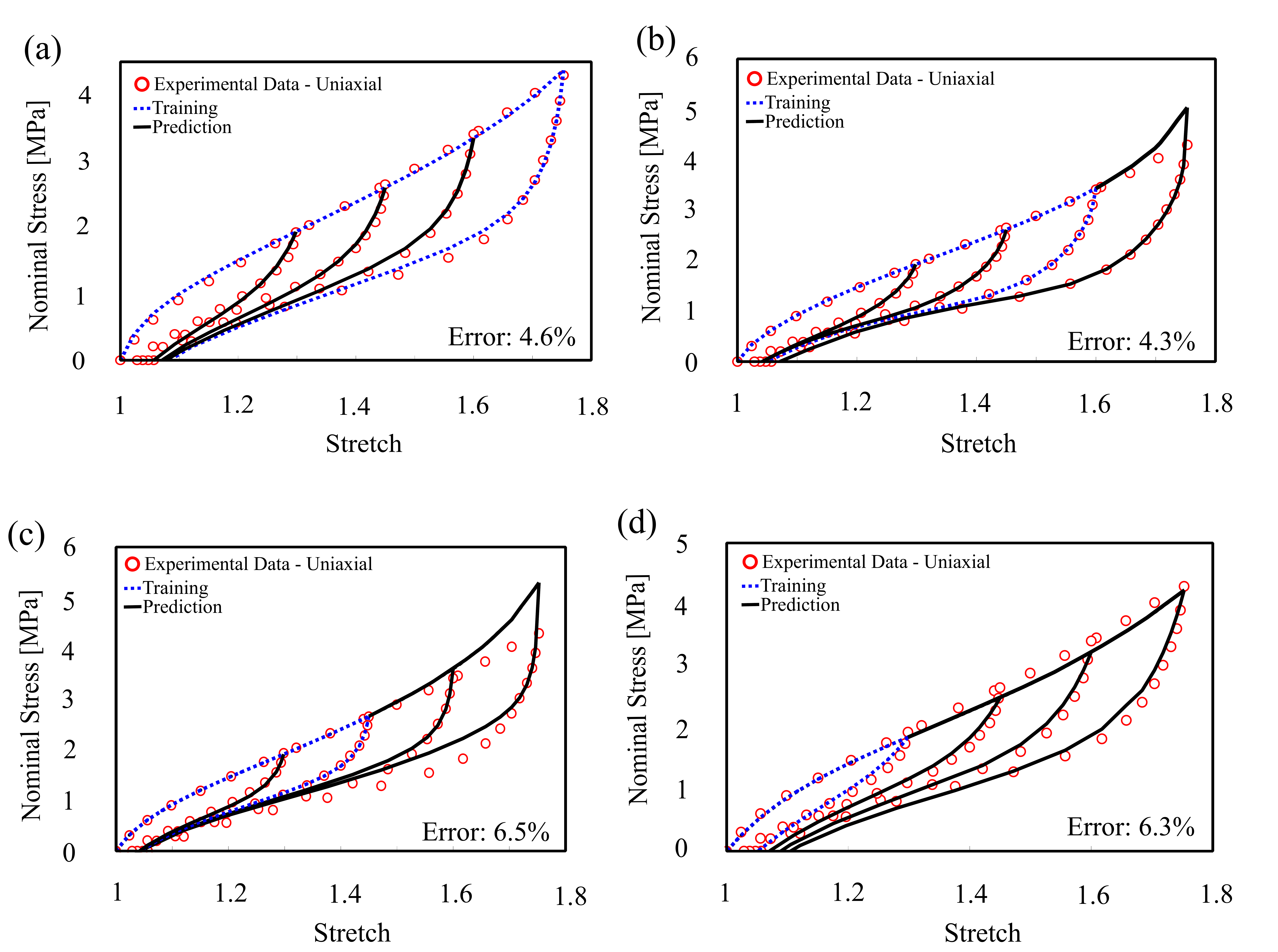}}
\caption{Model training and prediction of cyclic uniaxial tension with step-wise increasing of amplitude (Itskov's data set \cite{itskov2016rubber})}
\label{uni2}
\end{figure}

\begin{figure}[H]
\centerline{\includegraphics[width=1.1\textwidth]{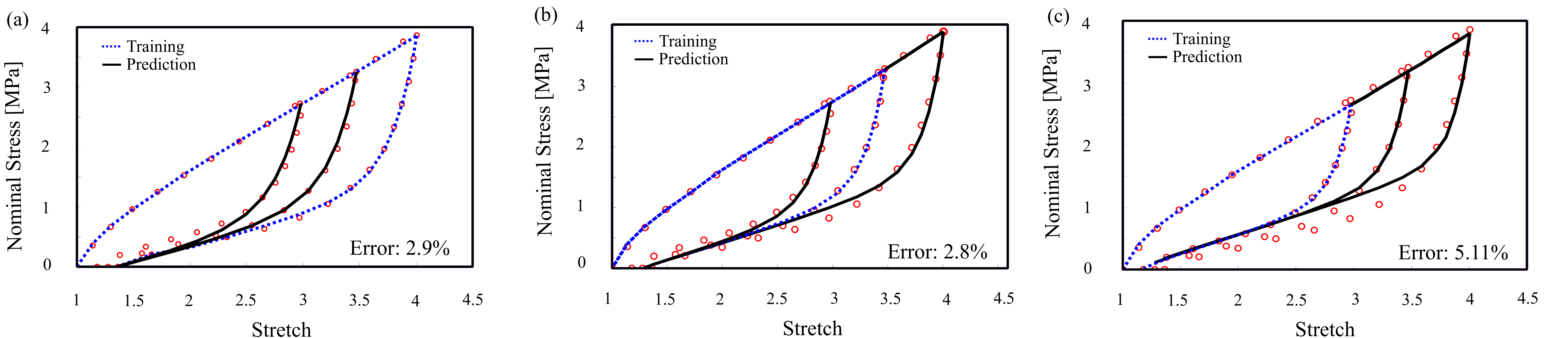}}
\caption{Model training and prediction of cyclic uniaxial tension with step-wise increasing of amplitude (data set \cite{zhong2019physically})}
\label{uni3}
\end{figure}

\paragraph{Convergency outside of the confidence interval} 
To investigate the convergency of the proposed model outside of the confidence, prediction of the inelastic behaviour on stretch amplitudes larger than the confidence interval were illustrated on two different sets of experimental data on rubber, Itskov's \cite{itskov2016rubber} and Zhong's \cite{zhong2019physically} dataset.
While we strongly recommend the users to stay within confidence interval, the model prediction accuracy outside of the confidence interval shows the relevance and reliability of the model in extreme cases which is mainly resulted from the constraints induced by knowledge infused into the model. Results indicated that the trend and proposed model performance (Fig. \ref{uni2} and Fig. \ref{uni3}). Here, we gradually reduce the confidence interval by using smaller range of training data to see the drop in quality of predictions. As expected, despite accuracy reduction, there is no significant change in the model predictions profile which is not usually the case for extrapolation methods. {In Fig.} \ref{uni2}.a, {we trained the model with the largest amplitude. As we expected, the error in training and prediction is} $4.6\%$. {As we reduce the amplitude of training in Fig.} \ref{uni2}, {we see that the error has increased generally. There is an instability in the errors and overestimating in Fig.} \ref{uni2}.b and Fig. \ref{uni2}.c {which root from numerical simplification and choosing same neural network structures and activation functions for different sets of data for training. The important point is that we want to show that the model is accurate for different modes of training. To ensure our result is general for different elastomers, we did the same training procedure for another dataset (Fig.} \ref{uni3}). {The result shows the same results as the last dataset.}

\section{Conclusion}
\label{Con}
A physics-informed data-driven constitutive model for cross-linked polymers is developed by embedding Neural networks into a multi-scale model. We propose a systematic approach to reduce the order of the constitutive mapping by leveraging existing knowledge of polymer science, continuum physics, and statistical mechanics. We use our model to predict the mechanical behavior of filled elastomers. The results indicate that our model can easily capture multiple inelastic effects in the behavior of the materials, is significantly less data-dependent, has lower dimensionality, and is interpretable. To illustrate the superior performance of knowledge-driven models developed by this approach, its predictions are bench-marked against several experimental data sets. We compare the stress responses from Treloar's data set in our model with several well-known models to show the accuracy and simplicity of our model. {In summary, our model provides a hyper-elastic constitutive model which captures damage of polymer chains for cross-linked elastomers for quasi-static loading. In the future, the proposed model can be further extended to include the effect of the deformation rate. The modular platform nature of the proposed model allows the addition of such effects.}

\section*{Appendix}
\label{append}

\appendix

\section{Frame Independency}
\label{frame}
Frame objectivity, during rigid body motion, requires strain energy of the material remains unchanged. Thus, the material response should not depend on the choice of the reference frame. The strain energy frame independency can be written as

\begin{equation}
    \Psi_m(\tens Q \tens F) = \Psi_m (\tens F),
\end{equation}

where $\tens Q$ is the rotation tensor. So, a constitutive law is frame independent if energy is left rotationally invariant. The mentioned condition is satisfied when the strain energy is a function of the right Cauchy-Green deformation tensor $\tens C$, due to

\begin{equation}
    \tens{C}^{+} = (\tens{F}^{+})^T \tens{F}^{+}= \tens{F}^{T} \tens{Q}^{T} \tens{Q} \tens{F}= \tens{F}^{T} \tens{F}=\tens C,
\end{equation}
which $\tens{F}^{+}=\tens{F} \tens{Q}$. The proposed model is a function of right Cauchy-Green deformation tensor. So, the frame independency condition is satisfied automatically.

\section{Thermodynamic Consistency}

\subsection{Polyconvexity}
\label{poly}

Polyconvexity is one of the known conditions which ensure the thermodynamic consistency. In this section, we briefly describe sufficient but not necessary free energy function conditions which guarantee the existence of minimizers of some variational principles. In order to understand polyconvexity, we start with some properties of convexity. Consider that $\Psi_m\left(\tens{F}\right)$ is the strain energy function on set of $K$. We can say $\Psi_m\left(\tens{F}\right)$ is convex on set of $K$ if hessian matrix of $\Psi_m\left(\tens{F}\right)$ be positive in that set.

\begin{equation}
    \label{convexity}
    {D}^{2}{\Psi_m\left(\tens{F}\right)}.\left({H},{H}\right) \geqslant {0},
\end{equation}

and for proof of polyconvexity we can mention that ${\tens{F}}\to {\Psi_m \left(\tens{F}\right)}$ is polyconvex if and only if there exist a function $G$ such that

\begin{equation}
    \label{polyconvexity}
    {\Psi_m\left(\tens{F}\right)} = {G}\left(\tens{F}, \adj \tens{F}, \det \tens{F}\right),
\end{equation}

and the function $G$ is convex. Besides, $\adj \tens{F} = \frac{\tens{F}^{-1}}{\det \tens{F}} $ and the implication chain shows relations from convexity to ellipticity.
\begin{center}
    convexity $\to$ polyconvexity $\to$ quasiconvexity $\to$ ellipticity
\end{center}

The Hessian matrix of the strain energy is positive if

\begin{equation}
    \frac{{\partial{\Psi_m}}^2}{ {{\partial}^2 {{{{\lambda_j}}}^{{\vect{d}_i}}}}}= {\sum_{i = 1}^{N_d} {w_i} {\frac{{{\partial {{\psi_{j}}^{ \vect{d_{i}}}}}}^2}{{{\partial}^2 {{{\lambda_j}}^{{\vect{d}_i}}}}}} }= {\sum_{i = 1}^{N_d} {w_i} {\frac{\partial {ANN_{j}(\tens{W_{j}},{\lambda_j}^{\vec{d_{i}}}, {{\lambda_j}_{max}}^{\vec{d_{i}}})}}{{\partial{{{\lambda_j}}^{{\vect{d}_i}}}}}} } > 0,  \quad \quad for \text{ } j=1,2,..., N_s,
\end{equation}

If weights which connect the input of $\lambda_j$ to other neurons be positive, the proposed model holds the condition of polyconvexity.     
    
\subsection{Second Law of Thermodynamic}
\label{thermo}
Because all of the constitutive models should satisfy the second law of thermodynamic, the satisfaction of this law should be checked for the proposed model. On the other hand, checking Clausius-Duhem inequality would be enough for this.
Because ${\lambda_j}_{max}$ is internal variables in the strain energy function of Cross-linked polymers, we can reduce the second law of thermodynamics to Clausius-Duhem inequality that shows thermodynamic consistency of the model in direction $d_{i}$. This inequality can be written as

\begin{equation}
    \label{thermo1}
    \frac{\partial {\Psi_m}}{\partial {{\lambda_{jmax}}^{{\vect{d}_i}}}} \leq 0 \quad \quad \quad \quad \forall \text{ } \vect{d}   \quad \quad \quad \quad for \text{ } j=1,2,...,N_s,
\end{equation}

If we consider the energy of matrix as 

\begin{equation}
    \label{energy}
    \Psi_m  = \sum_{i = 1}^{N_d}  \sum_{j = 1}^{N_s} \left({\mathop {{\psi_{j}}^{ \vect{d_{i}}}}}\right) \mathop {w_i },
\end{equation}
which

\begin{equation}
    \label{single}
    \mathop {\psi_{j}}^{ \vect{d}}=ANN_{j}(\tens{W_{j}},{\lambda_j}^{\vec{d_{i}}}, {{\lambda_j}_{max}}^{\vec{d_{i}}}), 
\end{equation}
thus, Clausius-Duhem can be written as

\begin{equation}
    \frac{\partial{\Psi_m}}{ {\partial {{{\lambda_j}_{max}}^{{\vect{d}_i}}}}}= {\sum_{i = 1}^{N_d} {w_i} {\frac{\partial {{\psi_{j}}^{ \vect{d_{i}}}}}{{\partial {{{\lambda_j}_{max}}^{{\vect{d}_i}}}}}} }= {\sum_{i = 1}^{N_d} {w_i} {\frac{\partial {ANN_{j}(\tens{W_{j}},{\lambda_j}^{\vec{d_{i}}}, {{\lambda_j}_{max}}^{\vec{d_{i}}})}}{{\partial{{{\lambda_j}_{max}}^{{\vect{d}_i}}}}}} } \leq 0, \quad \quad \quad \quad for \text{ } j=1,2,...,N_s,
\end{equation}

If weights that connect the input of ${\lambda_j}_{max}$ to other neurons be negative, the proposed model holds the condition of thermodynamic consistency.

\section{Integration Point of Micro-Sphere Approach}

\begin{table}[H]
\centering
\small
\caption{Integration points and weighting factors of the unit-sphere}
\begin{tabular}{ccccc}
\hline
\multicolumn{1}{c}{i}  & \multicolumn{1}{c}{${d_{i}}(1)$}             & \multicolumn{1}{c}{${d_{i}}(2)$}              & \multicolumn{1}{c}{${d_{i}}(3)$}             & \multicolumn{1}{c}{$w_{i}$}               \\ \hline
1                        & 0.0                                 & 0.0                                  & 1.0                                 & 0.0265214244093                      \\
2                        & 0.0                                 & 1.0                                  & 0.0                                 & 0.0265214244093                      \\
3                        & 1.0                                 & 0.0                                  & 0.0                                 & 0.0265214244093                      \\
4                        & 0.0                                 & 0.707106781187                       & 0.707106781187                      & 0.0199301476312                      \\
5                        & 0.0                                 & 0.707106781187                       & 0.707106781187                      & 0.0199301476312                      \\
6                        & 0.707106781187                      & 0.0                                  & 0.707106781187                      & 0.0199301476312                      \\
7                        & 0.707106781187                      & 0.0                                  & 0.707106781187                      & 0.0199301476312                      \\
8                        & 0.707106781187                      & 0.707106781187                       & 0.0                                 & 0.0199301476312                      \\
9                        & 0.707106781187                      & 0.707106781187                       & 0.0                                 & 0.0199301476312                      \\
10                       & 0.836095596749                      & 0.387907304067                       & 0.387907304067                      & 0.0250712367487                      \\
11                       & 0.836095596749                      & 0.387907304067                       & 0.387907304067                      & 0.0250712367487                      \\
12                       & 0.836095596749                      & 0.387907304067                       & 0.387907304067                      & 0.0250712367487                      \\
13                       & 0.836095596749                      & 0.387907304067                       & 0.387907304067                      & 0.0250712367487                      \\
14                       & 0.387907304067                      & 0.836095596749                       & 0.387907304067                      & 0.0250712367487                      \\
15                       & 0.387907304067                      & 0.836095596749                       & 0.387907304067                      & 0.0250712367487                      \\
16                       & 0.387907304067                      & 0.836095596749                       & 0.387907304067                      & 0.0250712367487                      \\
17                       & 0.387907304067                      & 0.836095596749                       & 0.387907304067                      & 0.0250712367487                      \\
18                       & 0.387907304067                      & 0.387907304067                       & 0.836095596749                      & 0.0250712367487                      \\
19                       & 0.387907304067                      & 0.387907304067                       & 0.836095596749                      & 0.0250712367487                      \\
20                       & 0.387907304067                      & 0.387907304067                       & 0.836095596749                      & 0.0250712367487                      \\ 
\multicolumn{1}{c}{21} & \multicolumn{1}{c}{0.387907304067} & \multicolumn{1}{c}{-0.387907304067} & \multicolumn{1}{c}{0.836095596749} & \multicolumn{1}{c}{0.0250712367487} \\ \hline
\end{tabular}
\end{table}

\pagebreak

\bibliographystyle{file}
\bibliography{file.bib}

\end{document}